\documentclass[12pt]{article}
\usepackage[dvips]{color}
\usepackage{epsfig}
\usepackage{amsmath}
\usepackage{graphicx}

\begin{document}
\title{\bf Interacting Ghost Dark Energy Models in the Higher Dimensional Cosmology}
\author{{J. Sadeghi$^{a}$ \thanks{Email: pouriya@ipm.ir},\hspace{1mm} M. Khurshudyan$^{b}$ \thanks{Email: khurshudyan@yandex.ru},\hspace{1mm}
and H. Farahani$^{a}$ \thanks{Email:
h.farahani@umz.ac.ir}}\\
$^{a}${\small {\em Department of Physics, University of Mazandaran, Babolsar, Iran}}\\
{\small {\em P .O .Box 47416-95447, Babolsar, Iran}}\\
$^{b}${\small {\em Institute for physical research, National Academy of Sciences of Armenia, Ashtarak, Armenia}}} \maketitle
\begin{abstract}
We investigate interacting ghost dark energy models in higher dimensional cosmology. We attempt to model dark matter within a barotropic fluid with $P_{b}=\omega(t)_{b}\rho$. In this work we consider four different models based on choosing equation of state parameter and interaction term. We confirm that our models agree with observational data.
\end{abstract}
\section{Introduction}
In modern cosmology it is observed, by using high redshift type Ia supernovae [1-3], that the Universe expand with acceleration. Then, several
observations show that the density of matter is very much less than
critical density [4], observations of Cosmic Microwave Background
(CMB) anisotropy indicate that the Universe is flat and the total
energy density is very close to the critical $\Omega_{\small{tot}}
\simeq1$ [5]. Faced with these results we started to find realistic
models to explain experimental data concerning to the nature of the
accelerated expansion of the Universe and a huge number of
hypothesis were proposed. For instance, in general relativity
framework, the desirable result could be achieved by so-called dark
energy, with
negative pressure and negative EoS parameter ($\omega<0$). The simplest model for a dark energy is a
cosmological constant introduced by Einstein.
This model has fine-tuning and coincidence problems. Absence of a fundamental mechanism
which sets the cosmological constant zero or very small value makes
researchers to go deeper and deeper in theories to understand the
solution of these problems. In order to alleviate these problems alternative models of dark energy suggest a
dynamical form of dark energy, which at least in an effective level,
can originate from a variable cosmological constant [6, 7], or from
various fields, such as a quintessence [8-10], a phantom field [11-15], or the combination of
quintessence and phantom in a unified model named quintom [16-20].
By using some basic of quantum gravitational principles one can
formulate several other models for dark energy, and in literature
they are known as holographic dark energy models
[21-25] and agegraphic dark energy models [26-28].\\
Interaction between components is proved to be other way which can solve coincidence problem. From observations no piece of evidence has been
so far presented against interactions between dark energy and dark matter.
From theoretical side we have not any known symmetry which prevents or suppresses a non-minimal coupling between dark energy and dark matter.\\
Research in theoretical cosmology proposes two possible ways to
explain later time accelerated expansion of the Universe. Remember
that field equations make connection between geometry and matter
content of Universe in a simple way. Therefore, there is two
possibilities either we should modify matter content which is coded
in energy-stress tensor or we should modify geometrical part
including different functions of Ricci scalar etc. Different type of
couplings between geometry and matter could give desirable effects
as well. Recently, several authors try to make connection between scalar field and other models of
dark energy. In literature, often used idea of fluid despite to
other ideas, because over the years we learned that modifications of
geometrical part of field equations can be codded in fluid
expression. Chaplygin gas and its extensions [29-40] are interesting models based of the above idea. This model is more appropriate
choice to have constant negative pressure at low energy density and
high pressure at high energy density.\\
Among various models of dark energy, a new model is called Veneziano ghost dark energy, has attracted a lot of interests in recent years
[41-47]. Indeed, the contribution of the ghosts field to the vacuum
energy in curved space or time-dependent backgrounds can regarded
as a possible candidate for the dark energy. Veneziano ghost is unphysical in
the QFT formulation in Minkowski space-time, but exhibits important
non trivial physical effects in the expanding Universe. In the recent work we considered interacting Ghost dark energy models with variable $G$ and $\Lambda$ [48]. Now, we would like to extend this work to the case including extra dimensions [49, 50, 51]. Extra dimensions introduced some times to obtain unified theory, for example in the Kaluza-Klein theory to explain
the unification of fundamental force with gravity. The extra dimensions also obtained in superstring theory via mathematical formulation. In the
last decade it was suggested that the extra dimensions may explain acceleration of the
universe. In the Ref. [50] a five dimensional FRW cosmology with static extra
dimension considered and field equations including cosmological constant obtained.

\section{Models and Field equations}
Here we would like to consider a Universe including dark matter and dark energy.
In this work, the dark matter will modeled as a barotropic fluid. Two different models for EoS parameter of the barotropic fluid will be considered as the following. In the first model we assume,
\begin{equation}\label{s1}
\omega(t)_{b}=\omega_{0}+\omega_{1}t\frac{\dot{H}}{H},
\end{equation}
and the second model considered the following EoS parameter,
\begin{equation}\label{s2}
\omega(t)_{b}=\omega_{0}\cos(tH)+\omega_{1}t\frac{\dot{H}}{H},
\end{equation}
where $\omega_{0}$ and $\omega_{1}$ are positive constants.\\
Dark energy will modeled within a Ghost dark energy with energy density,
\begin{equation}\label{eq:GDE}
\rho_{G}=\theta H,
\end{equation}
where $\theta$ is a constant. Therefore, the content of the Universe is an effective fluid with $\rho=\rho_{b}+\rho_{G}$ and $P=P_{b}+P_{G}$, giving EoS parameter as,
\begin{equation}\label{s4}
\omega_{tot}=\frac{P_{b}+P_{G}}{\rho_{b}+\rho_{G}}.
\end{equation}
This paper may be extension of the Ref. [48] to the case of higher dimensional space-time (opposite of that paper $G$ and $\Lambda$ are constant). This paper is organized as the following. In section 2 we recall field equations corresponding to higher dimension FRW Universe. Then in section 3 we introduce 4 different models and solve field equation numerically to obtain behavior of cosmological parameters. Finally in section 4 we summarize our results and give conclusion.\\

FRW metric with extra dimensions is represented by the following line element,
\begin{equation}\label{s5}
ds^{2}=ds^2_{FRW}+\sum_{i=1}^{d}{b(t)^{2}dx_{i}^{2}},
\end{equation}
where $d$ is the number of extra dimensions ($d=N-4$) and $ds_{FRW}^{2}$ represents the line element of the FRW metric in four dimensions which is given by,
\begin{equation}\label{s6}
ds^2_{FRW}=-dt^2+a(t)^2\left(dr^{2}+r^{2}d\Omega^{2}\right),
\end{equation}
where $d\Omega^{2}=d\theta^{2}+\sin^{2}\theta d\phi^{2}$, and $a(t)$
represents the scale factor. The $\theta$ and $\phi$ parameters are
the usual azimuthal and polar angles of spherical coordinates, with
$0\leq\theta\leq\pi$ and $0\leq\phi<2\pi$. The coordinates ($t, r,
\theta, \phi$) are called co-moving coordinates. $a(t)$ and $b(t)$ are the functions of $t$ alone represents the scale factors of 4-dimensional space time and extra dimensions respectively.\\
The field equations for the above non-vacuum higher dimensional space-time symmetry are as the following [49],
\begin{equation}\label{eq:Fridmman1}
3\frac{\dot{a}^{2}}{a^{2}}=\frac{d}{2}\frac{\ddot{b}}{b} +\frac{d^{2}-2d}{4}\frac{\dot{b}}{b^{2}}-\frac{d^{2}}{8}\frac{\dot{b}^{2}}{b^{2}}+\rho,
\end{equation}
\begin{equation}\label{eq:Fridmman2}
2\frac{\ddot{a}}{a}+\frac{\dot{a}^{2}}{a^{2}}=\frac{d}{2}\frac{\dot{a}}{a}\frac{\dot{b}}{b}+\frac{d^{2}}{8}\frac{\dot{b}^{}2}{b^{2}}-\frac{d}{8}\frac{\dot{b}^{2}}{b^{2}}-P,
\end{equation}
and,
\begin{equation}\label{eq:fridman3}
\frac{\ddot{b}}{b}+3\frac{\dot{a}}{a}\frac{\dot{b}}{b}=-\frac{d}{2}\frac{\dot{b}^{2}}{b^{2}}+\frac{\dot{b}^{2}}{b^{2}}-\frac{P}{2}.
\end{equation}
Energy conservation $T^{;j}_{ij}=0$ reads as,
\begin{equation}\label{eq:conservation}
\dot{\rho}+3H(\rho+P)=0,
\end{equation}
where the Hubble expansion parameter defined as,
\begin{equation}\label{eq:H}
H(t)=\frac{1}{d+3}(3\frac{\dot{a}}{a}+d\frac{\dot{b}}{b}).
\end{equation}
Well known fact is that the interaction between fluid components splits energy conservation equation, so we have,
\begin{equation}\label{eq:DE}
\dot{\rho}_{de}+(d+3)H(\rho_{de}+P_{de})=-Q,
\end{equation}
and,
\begin{equation}\label{eq:DM}
\dot{\rho}_{dm}+(d+3)H(\rho_{dm}+P_{dm})=Q.
\end{equation}
\section{Numerical results}
In this section we provide analysis of the model for a general case. We consider four different models based on choosing interaction term and EoS parameter. Numerically, we analyze about some cosmological parameters.
\subsection{Model 1}

\begin{figure}[h!]
 \begin{center}$
 \begin{array}{cccc}
\includegraphics[width=75 mm]{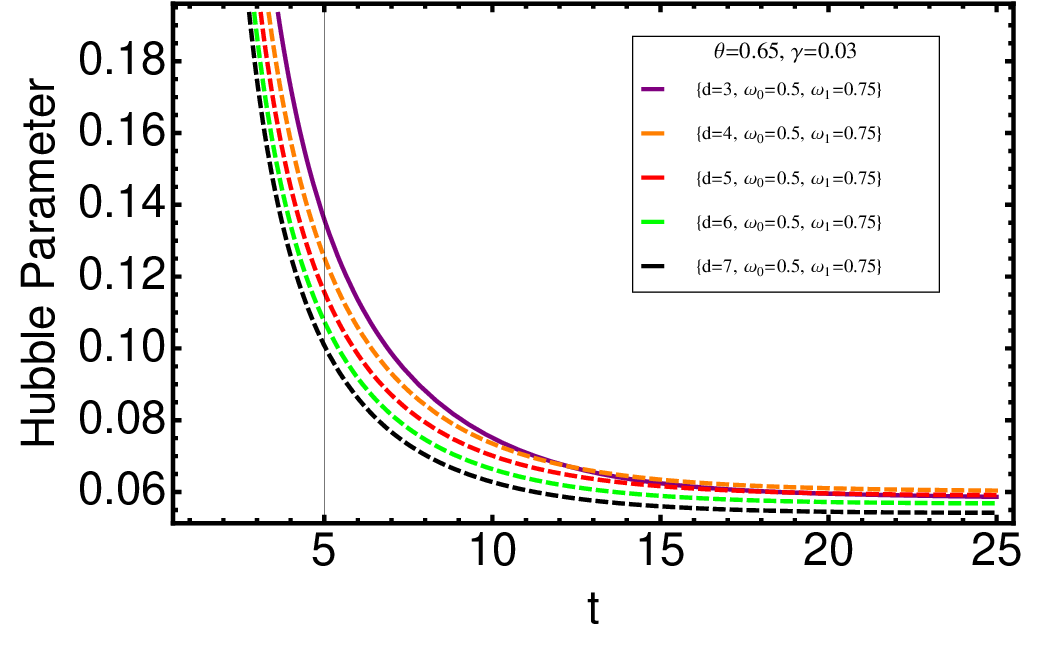}&\includegraphics[width=75 mm]{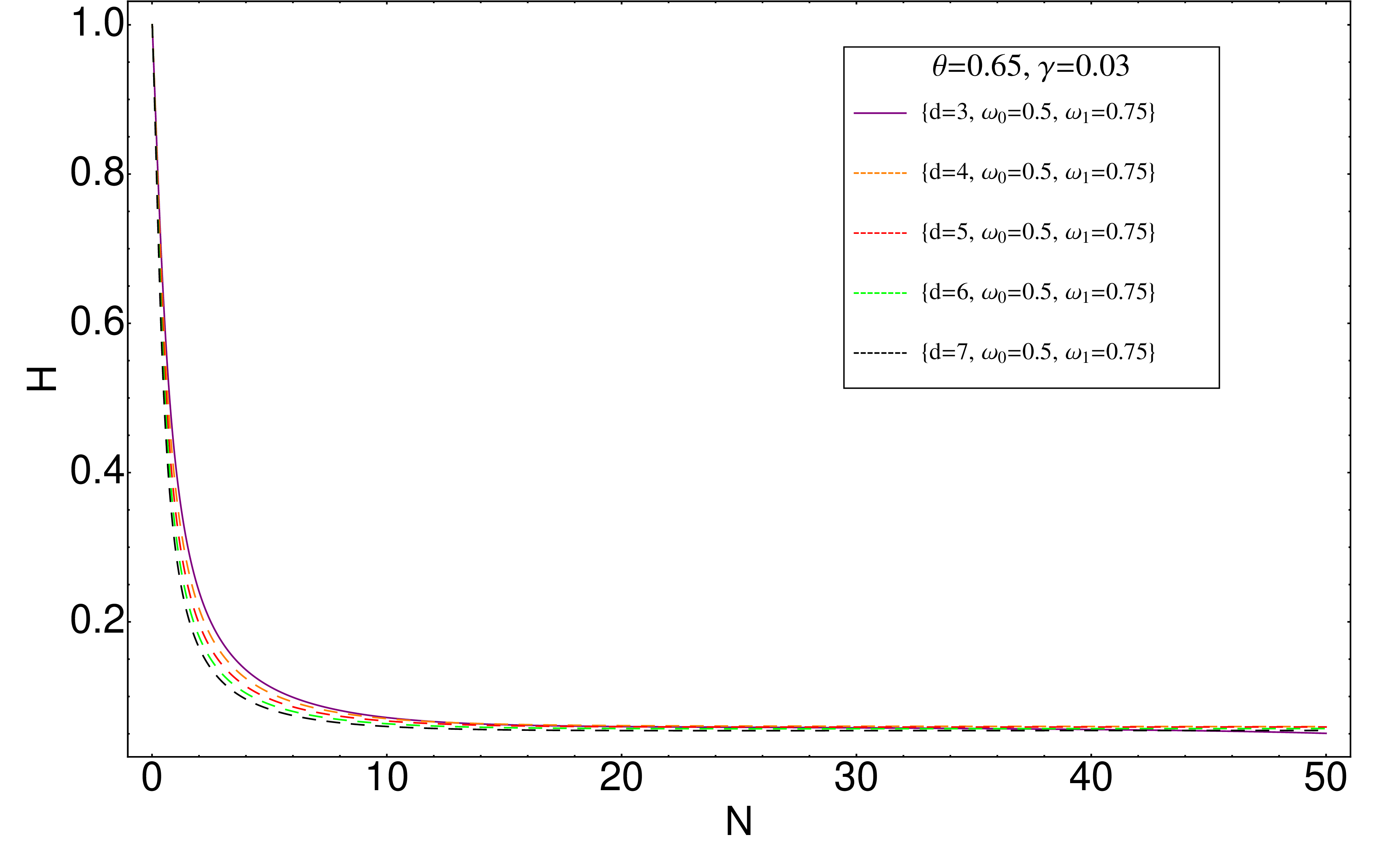}\\
\includegraphics[width=75 mm]{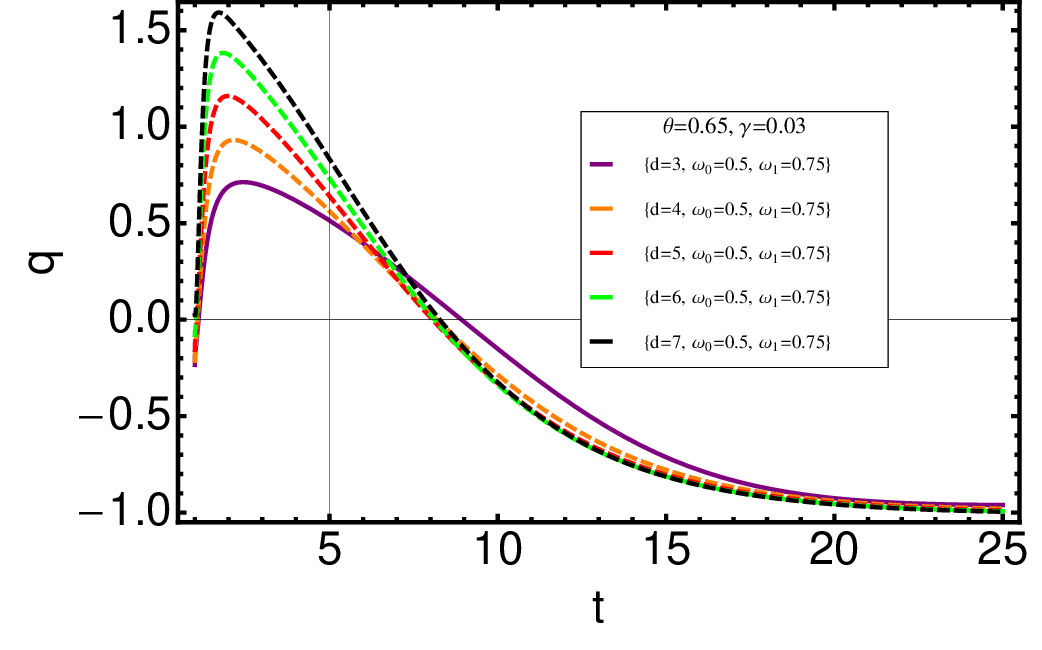}&\includegraphics[width=75 mm]{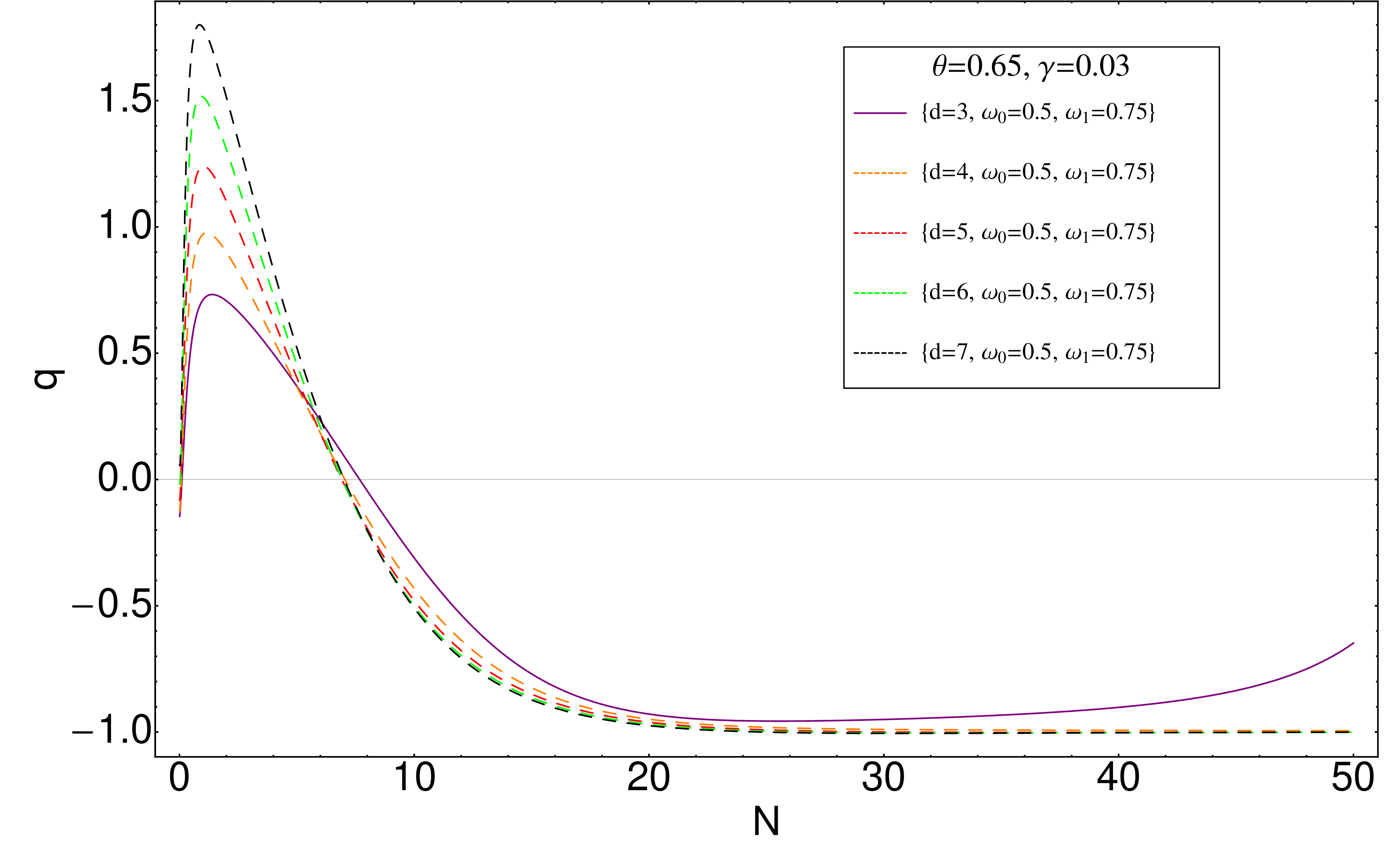}\\
\includegraphics[width=75 mm]{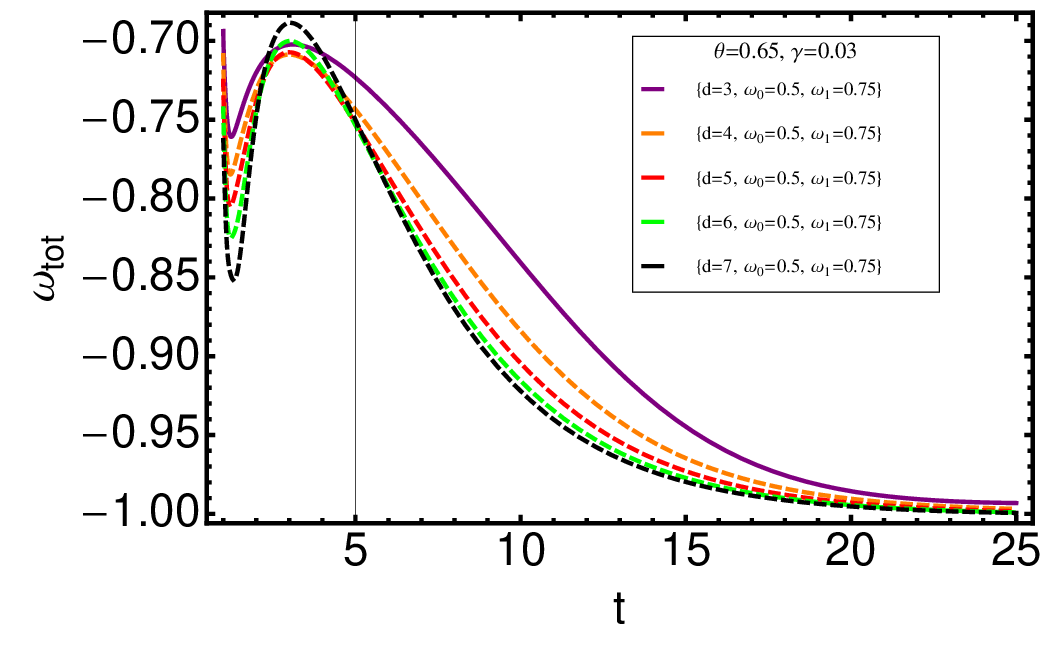}&\includegraphics[width=75 mm]{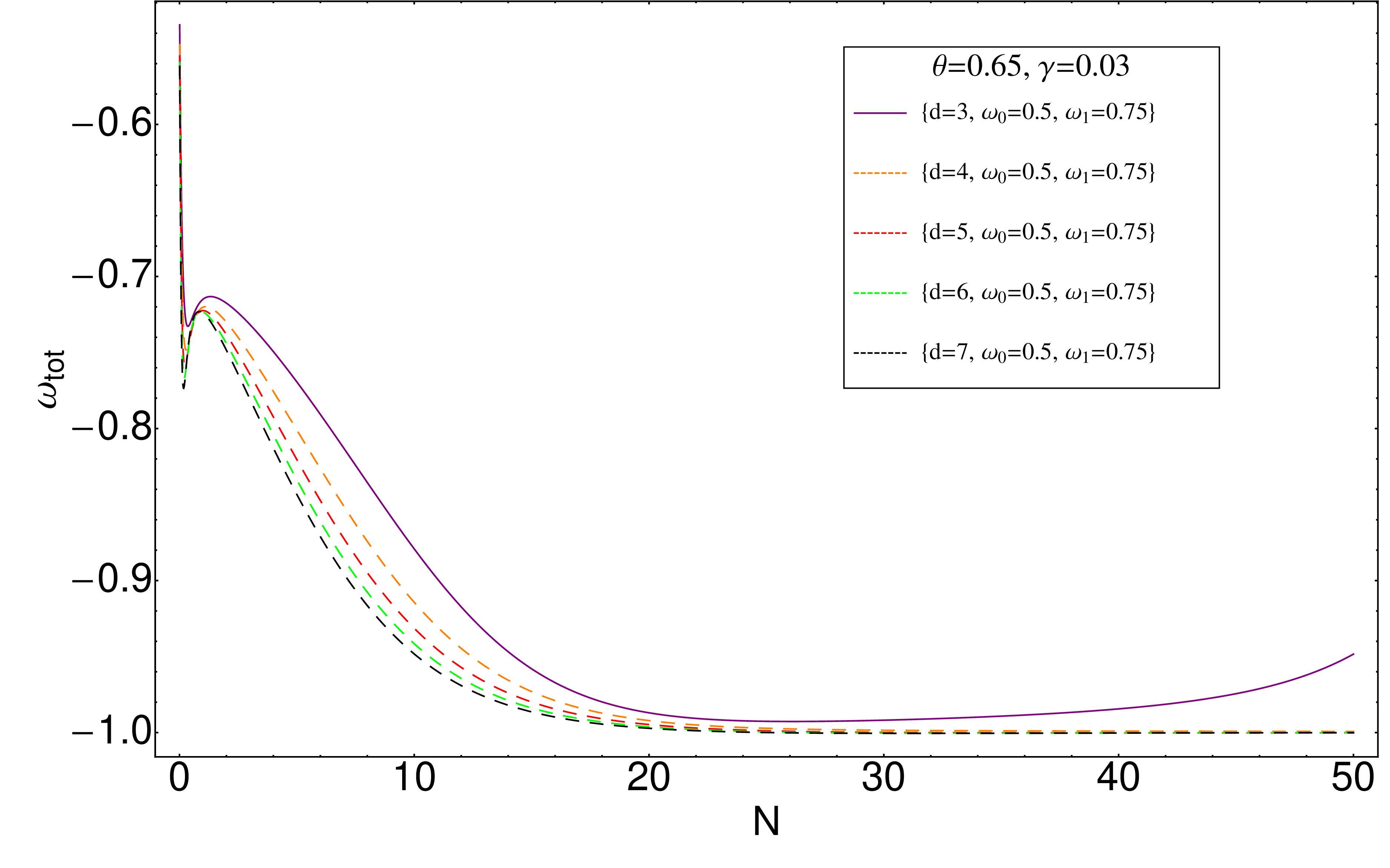}\\
 \end{array}$
 \end{center}
\caption{Behavior of $H$, $q$ and $\omega_{tot}$ against $t$ (left) and $N=\ln{a}$ (right). Model 1}
 \label{fig:1}
\end{figure}

Model 1 will characterize a Universe with interacting dark matter and dark energy, where dark matter is a barotropic fluid with,
\begin{equation}\label{s14}
\omega(t)_{b}=\omega_{0}+\omega_{1}t\frac{\dot{H}}{H},
\end{equation}
which already introduced by the equation (1).
Interaction $Q$ takes the form,
\begin{equation}\label{s15}
Q=(3+d)\gamma H \rho,
\end{equation}
where $\gamma$ is interaction coefficient and interpreted as strength of interaction. We expect $\gamma$ as small coefficient and exam different values to fix our results with observational data.
The dynamics of the energy density of the barotropic fluid according to (\ref{eq:DM}) can be written as,
\begin{equation}\label{s16}
\dot{\rho}_{b}+(3+d)(1-\gamma + \omega_{0}+\omega_{1}t\frac{\dot{H}}{H})H\rho_{b}=(3+d)\gamma H \rho_{G}.
\end{equation}
Concerning to the our assumption about the dark energy using (\ref{eq:DE}) we can obtain the pressure of dark energy as the following,
\begin{equation}\label{eq:PDE}
P_{G}=-\gamma \rho_{b}-(1+\gamma )\theta H -\frac{\theta \dot{H}}{(3+d)H}.
\end{equation}
Field equations with the last equations will allow us to obtain behavior of $H$, $q$ and $\omega_{tot}$. The graphical behavior of Hubble parameter, deceleration parameter and total EoS given in Fig. {\ref{fig:1}}. We see that Hubble parameter is a decreasing function. Deceleration parameter $q$ for the Universe of our consideration formally is given as,
\begin{equation}\label{s18}
q=-1-\frac{\dot{H}}{H},
\end{equation}
with $H$ as in (\ref{eq:H}).

From the Fig. 1 we can see that increasing $d$ decreases value of Hubble expansion parameter. Therefore, we can see that lower value of $d$ is more agree with current value of Hubble expansion parameter $H\sim70$ (0.07 in our scale). The second plot of the Fig. 1 represents deceleration parameter which yields to -1 in agreement with $\Lambda$CDM model. Also acceleration to deceleration phase transition illustrated in this figure. Finally the last plot of the Fig. 1 represent total EoS which is grater than -1 and shows quintessence like behavior of Universe. Hence, we can tell that the model 1 may agree with current observational data and will be acceptable.
\subsection{Model 2}

\begin{figure}[h!]
 \begin{center}$
 \begin{array}{cccc}
\includegraphics[width=75 mm]{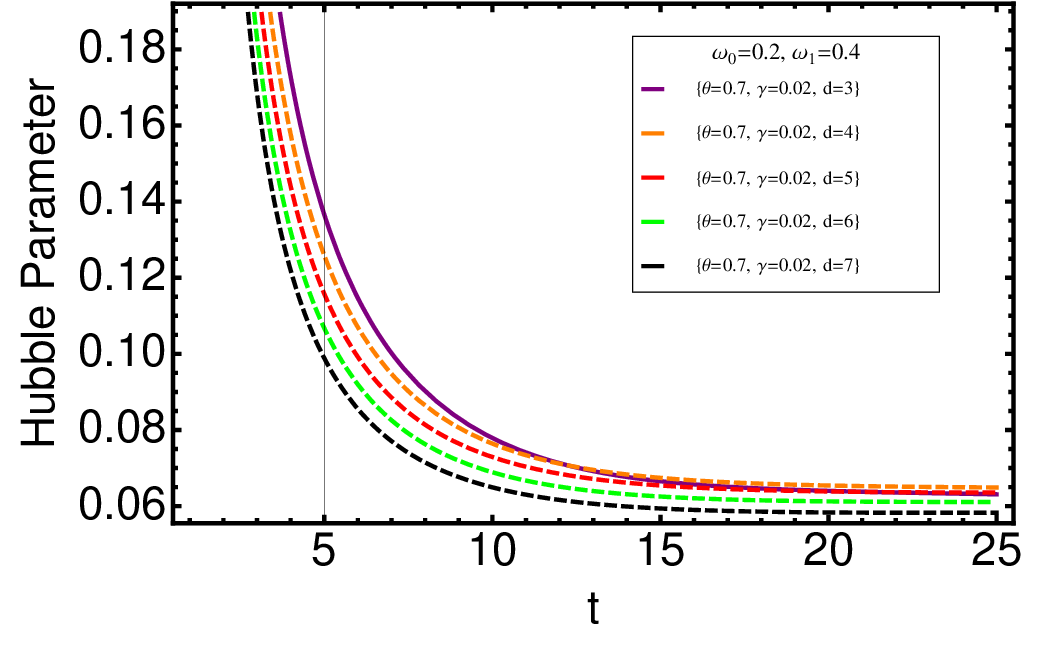} &
\includegraphics[width=75 mm]{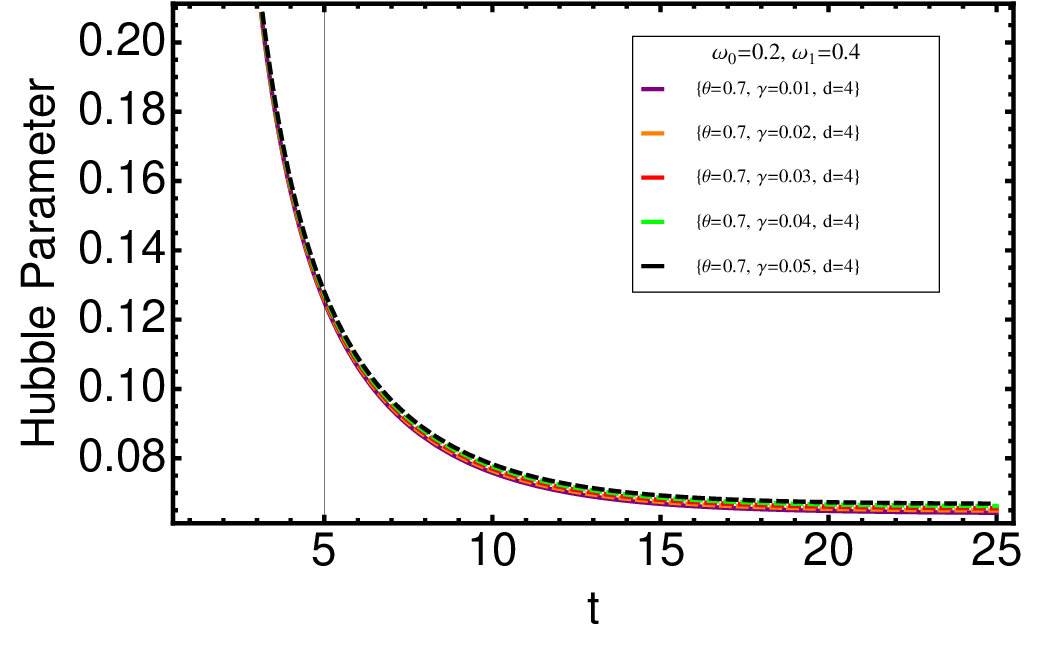}\\
\includegraphics[width=75 mm]{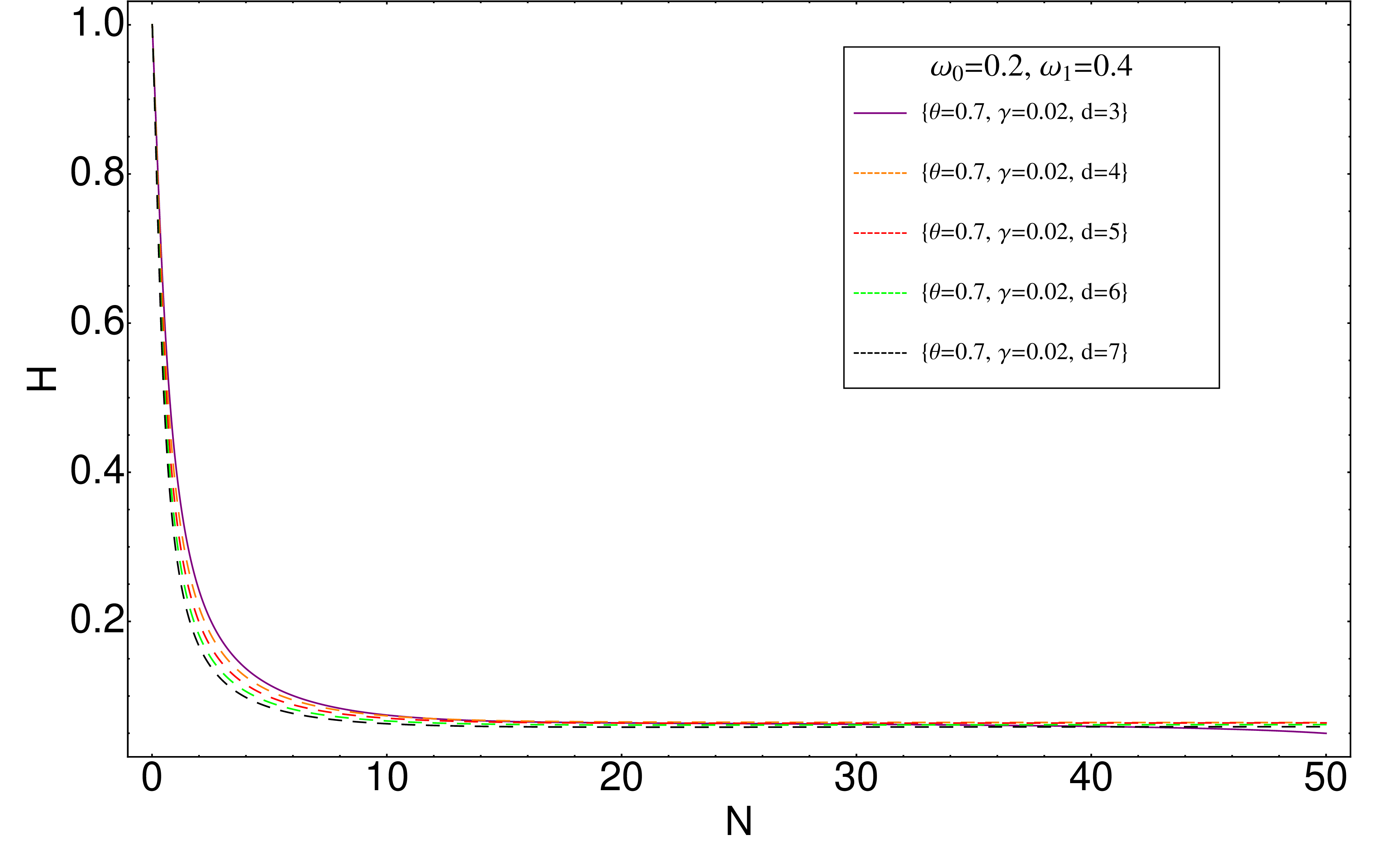} &
\includegraphics[width=75 mm]{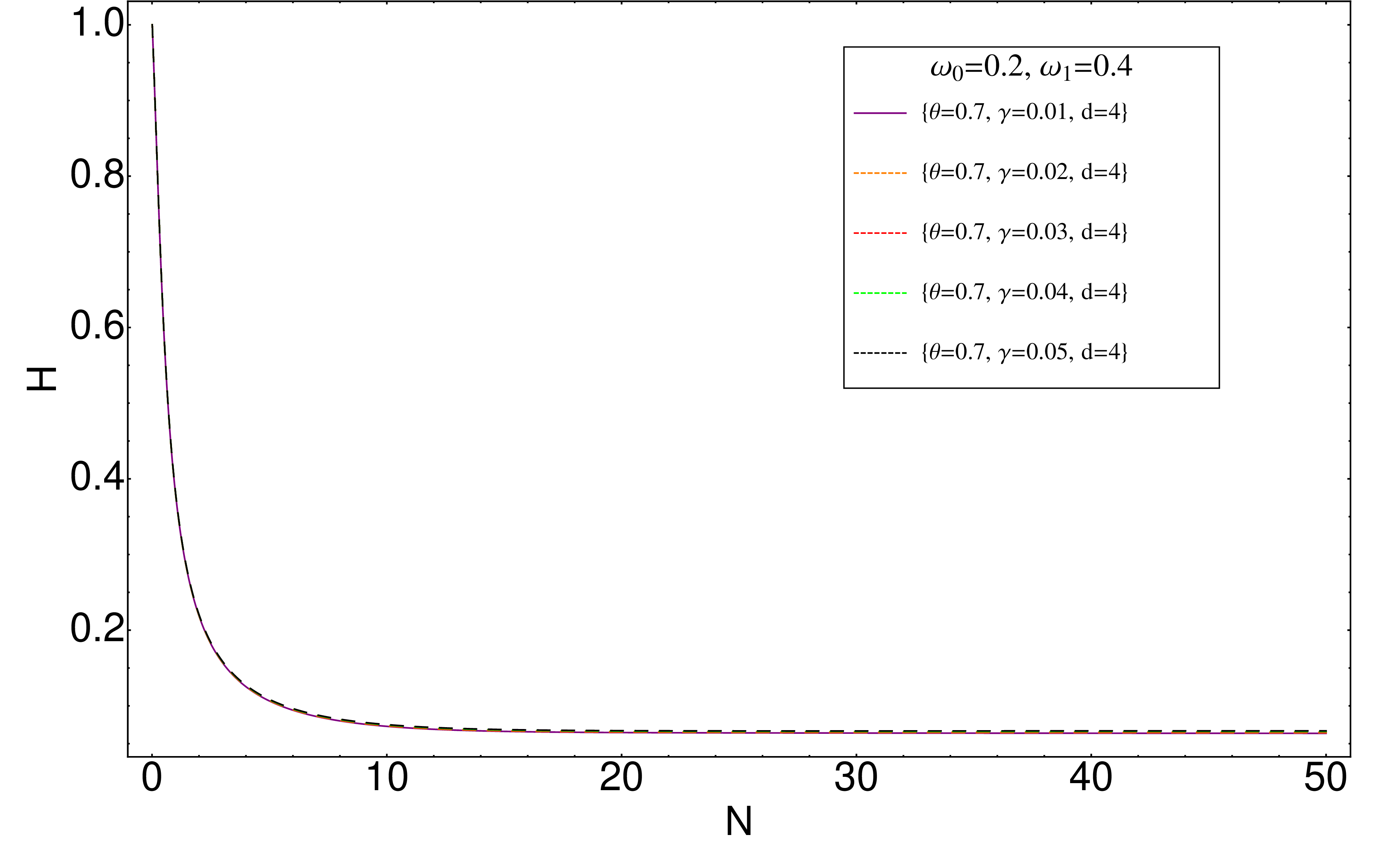}\\
 \end{array}$
 \end{center}
\caption{Behavior of $H$ against $t$  (up) and $N=\ln{a}$ (down). Model 2}
 \label{fig:2}
\end{figure}

\begin{figure}[h!]
 \begin{center}$
 \begin{array}{cccc}
\includegraphics[width=75 mm]{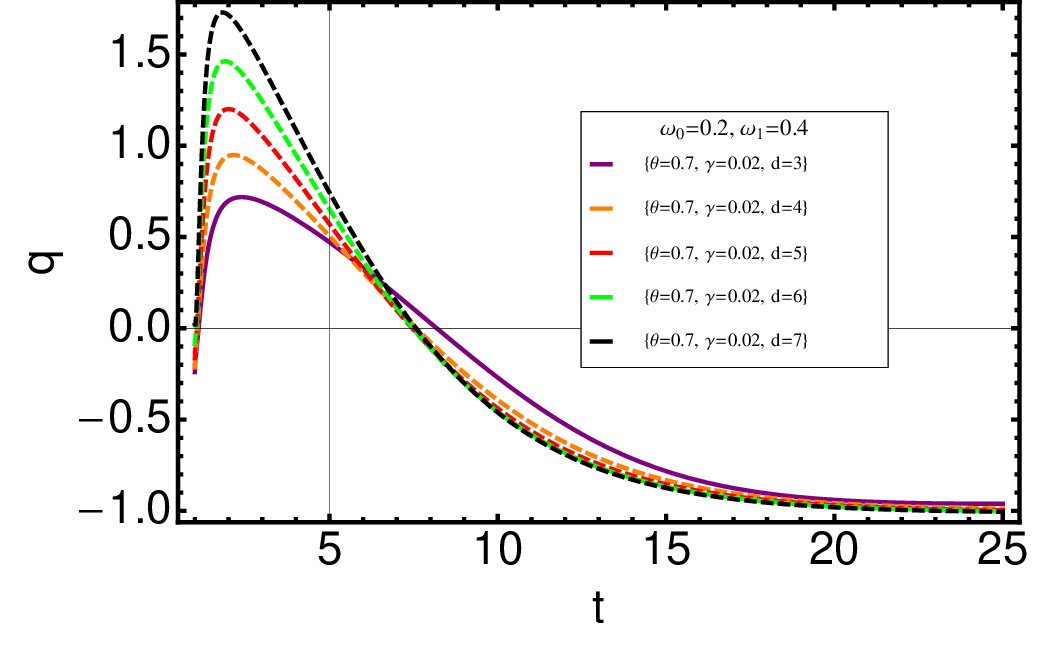} &
\includegraphics[width=75 mm]{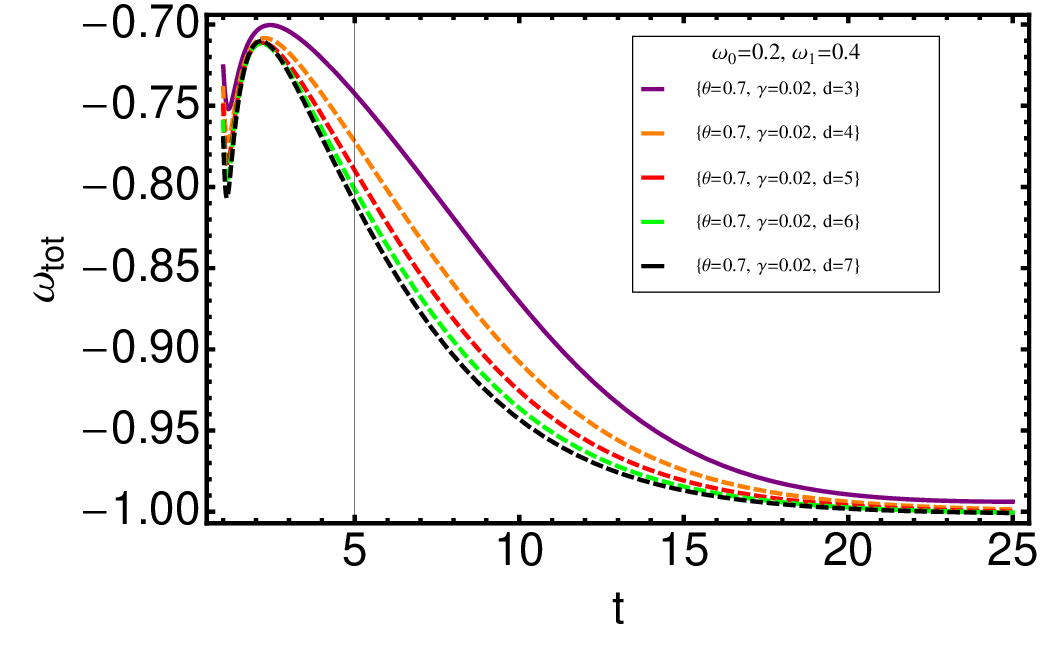}\\
\includegraphics[width=75 mm]{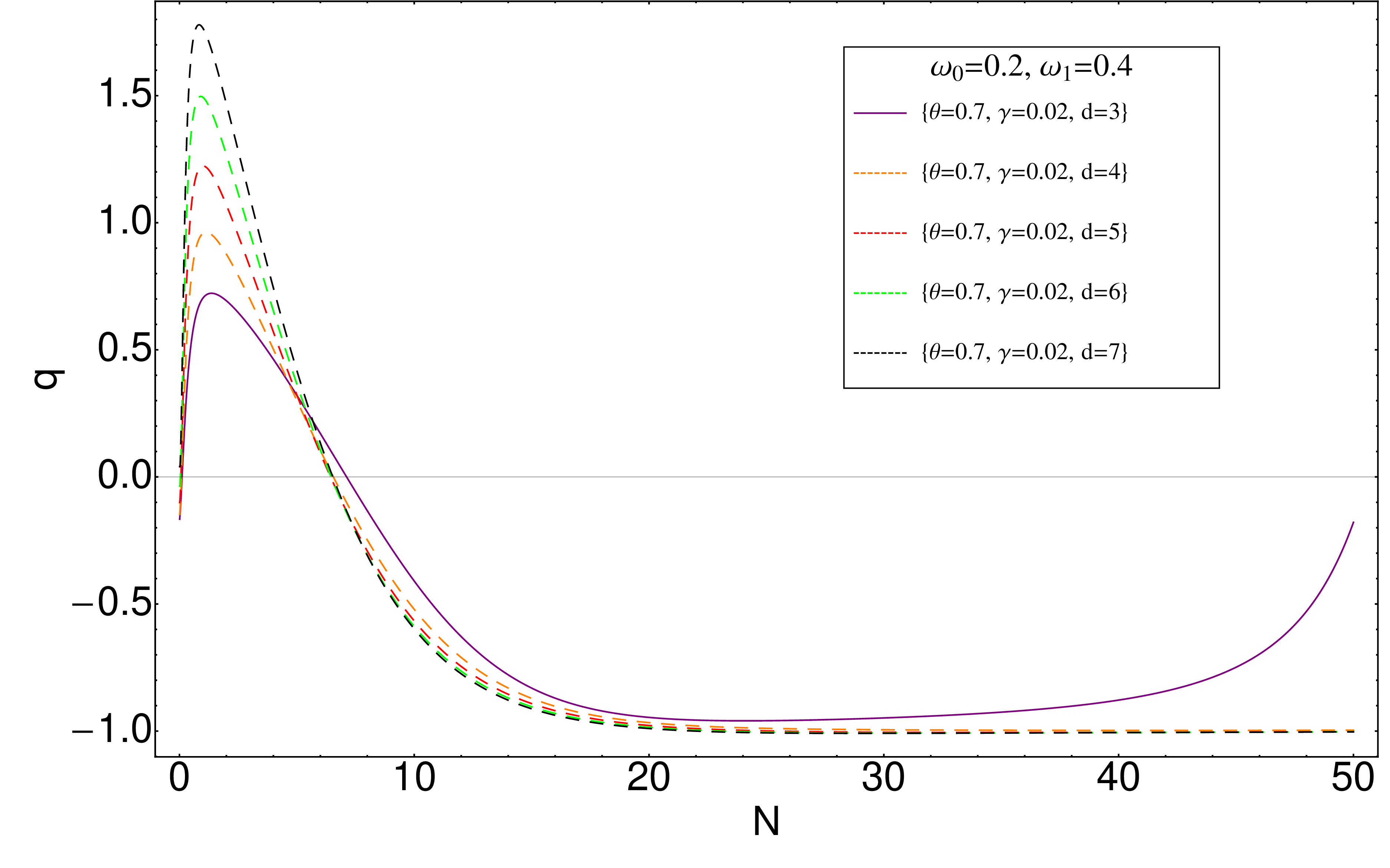} &
\includegraphics[width=75 mm]{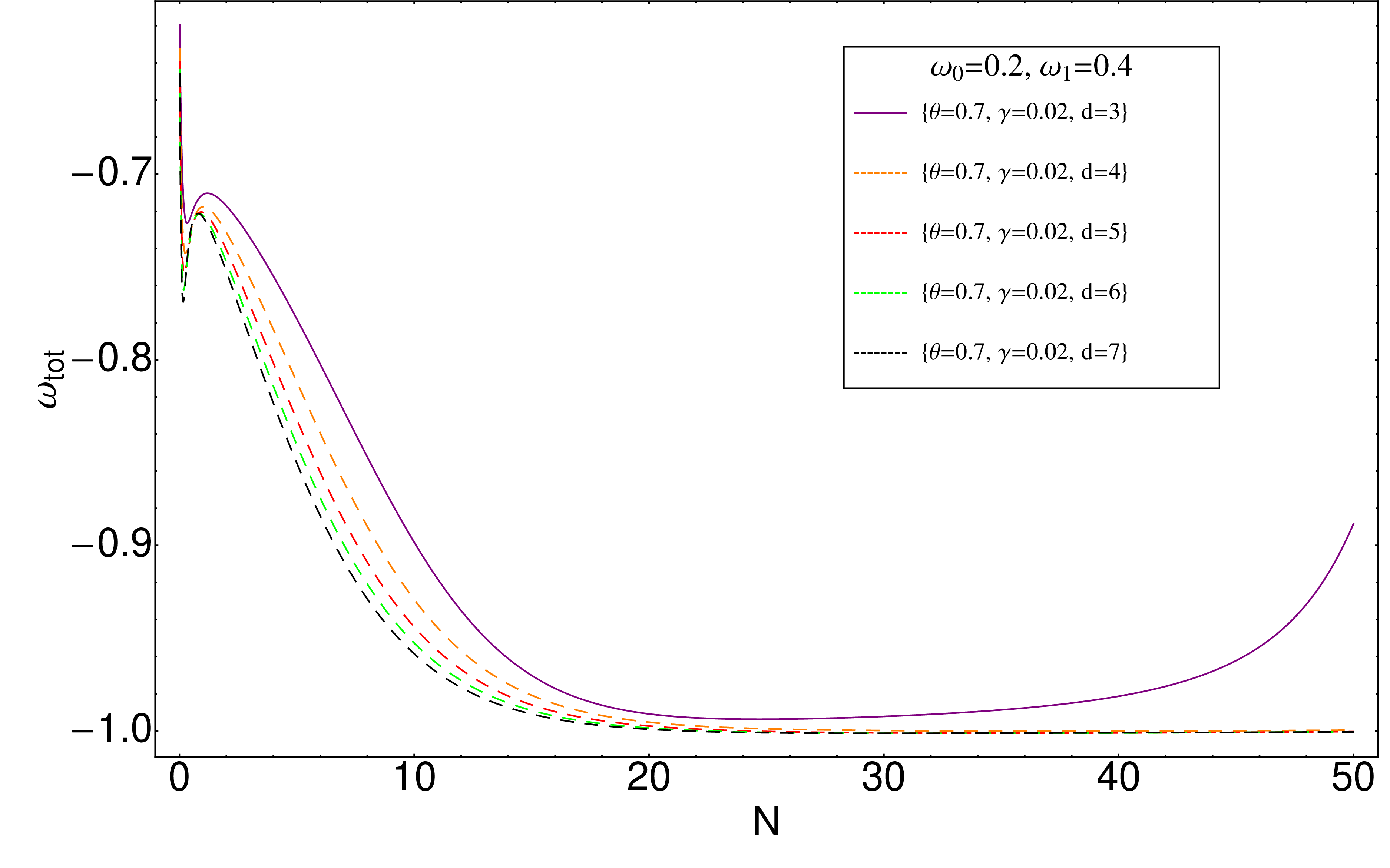}\\
 \end{array}$
 \end{center}
\caption{Behavior of $q$ and $\omega_{tot}$ against $t$ (up) and $N=\ln{a}$ (down). Model 2}
 \label{fig:3}
\end{figure}

In the second model of the Universe, dark matter is associated with a barotropic fluid with the following varying EoS parameter,
\begin{equation}\label{s19}
\omega(t)_{b}=\omega_{0}\cos(tH)+\omega_{1}t\frac{\dot{H}}{H}.
\end{equation}
Also, interaction $Q$ takes the form the equation (15).\\
For the dynamics of energy density $\rho_{b}$ we have a following differential equation,
\begin{equation}\label{eq:Dbf}
\dot{\rho}_{b}+(3+d)(1-\gamma+\omega_{0}\cos(tH)+\omega_{1}t\frac{\dot{H}}{H})H\rho_{b}=(3+d)\gamma H \rho_{G},
\end{equation}
which can be integrated easily giving $\rho_{b}(H)$. For this model the Hubble parameter $H$ is also a decreasing function.\\
From the left plot of the Fig. 2 we see that for the early stages of the evolution with increasing $d$ we will decrease the value of $H$, while for later stages of the evolution we see that with $d=4$ the current value of Hubble expansion parameter $H\sim70$ (0.07 in our scale) recovered. The right plot indicates the behavior of $H$ depends on interaction parameter $\gamma$. With increasing $\gamma$ we increase the value of $H$, when $\theta=0.65$, $\omega_{0}=0.5$, $\omega_{1}=0.75$ and $d=4$.\\
According to the Fig.(\ref{fig:3}) deceleration parameter $q$ indicates the possibility to have a Universe with 3 transitions in $q$. Starting from the evolution with $q<0$ for early stages of evolution, we have a transition to $q>0$ and finally there is a transition to $q<0$ at resent stages of evolution. Also, total EoS which is grater than -1 which shows quintessence like behavior of Universe.

\subsection{Model 3}

\begin{figure}[h!]
 \begin{center}$
 \begin{array}{cccc}
\includegraphics[width=75 mm]{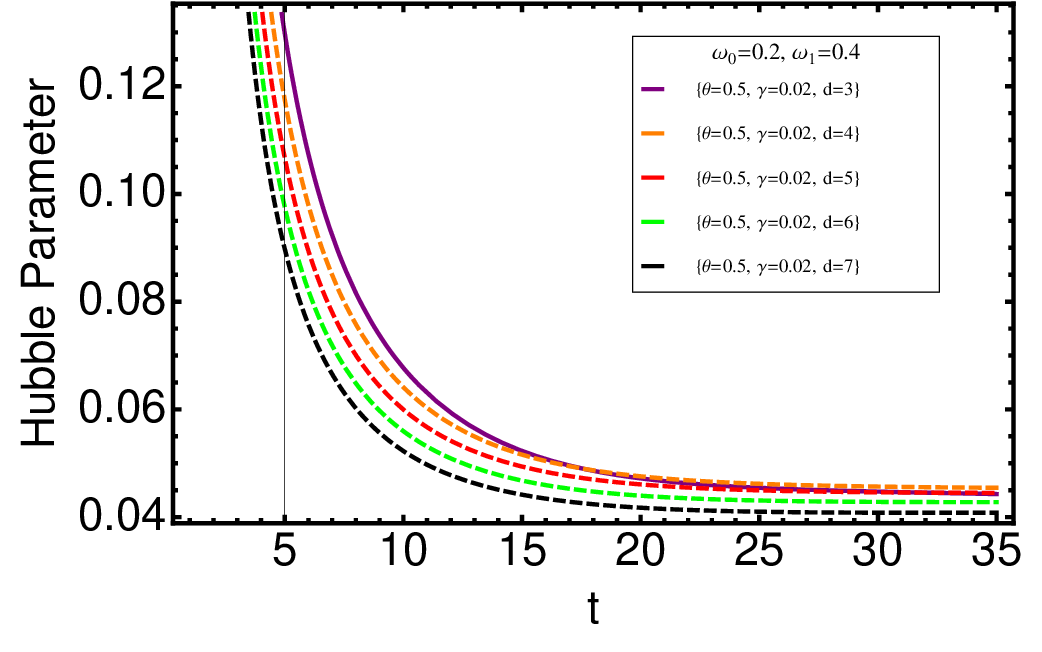}&\includegraphics[width=75 mm]{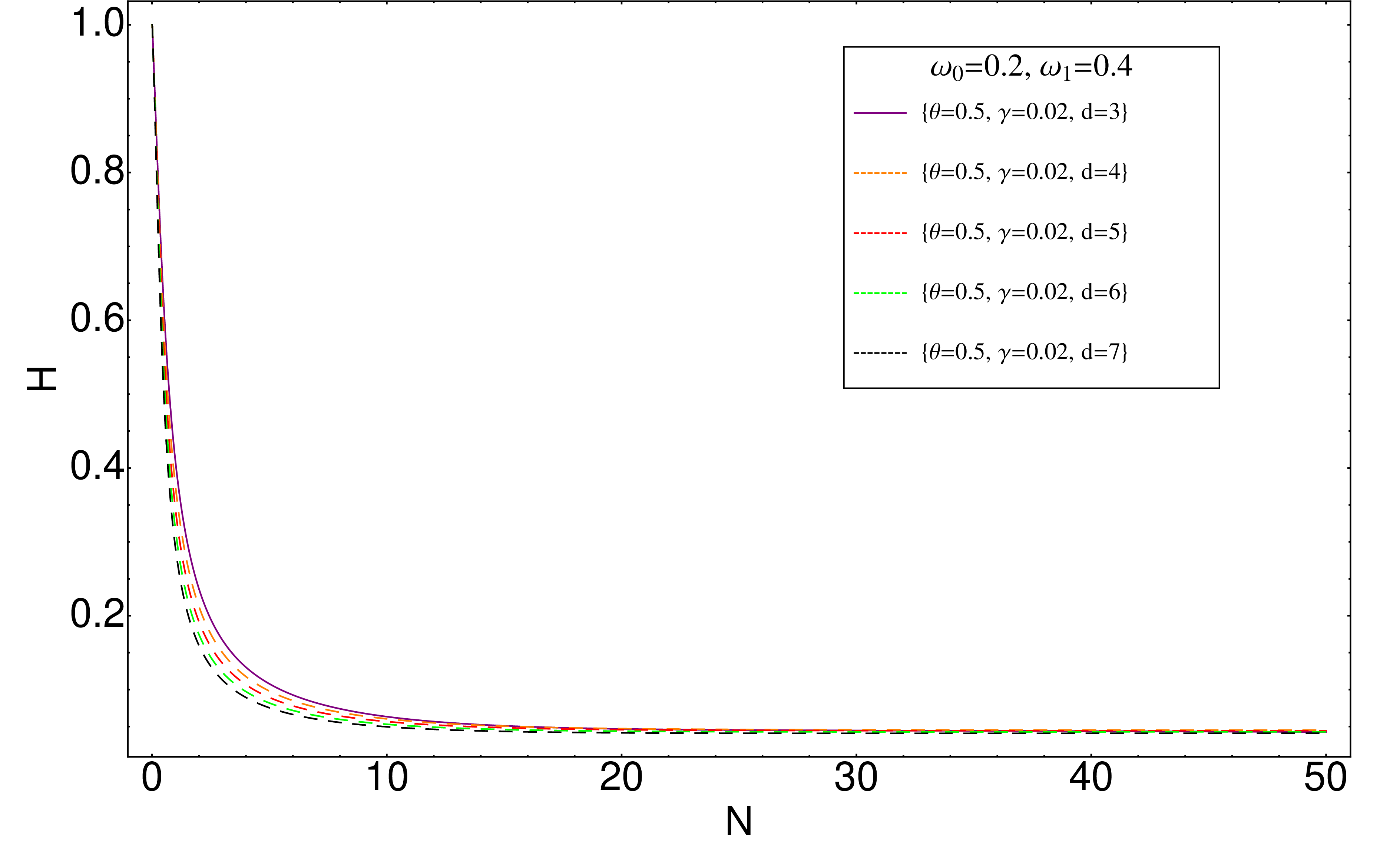}\\
\includegraphics[width=75 mm]{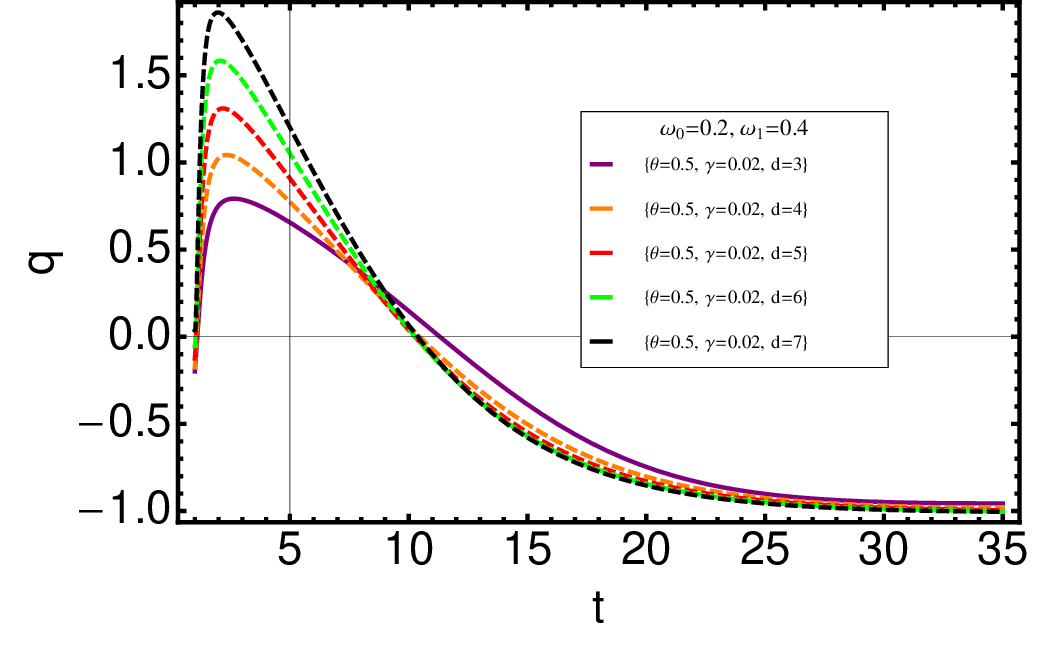}&\includegraphics[width=75 mm]{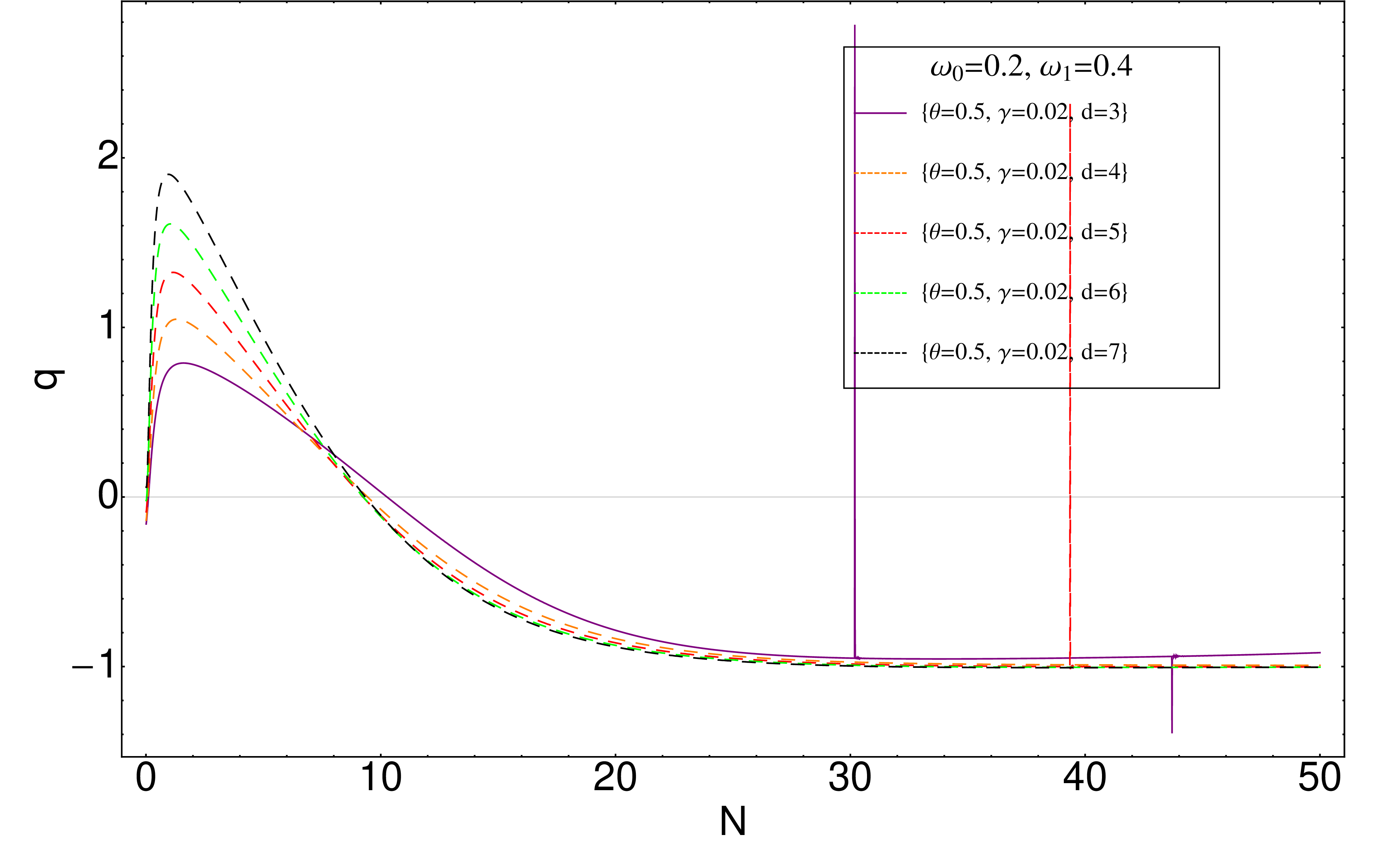}\\
\includegraphics[width=75 mm]{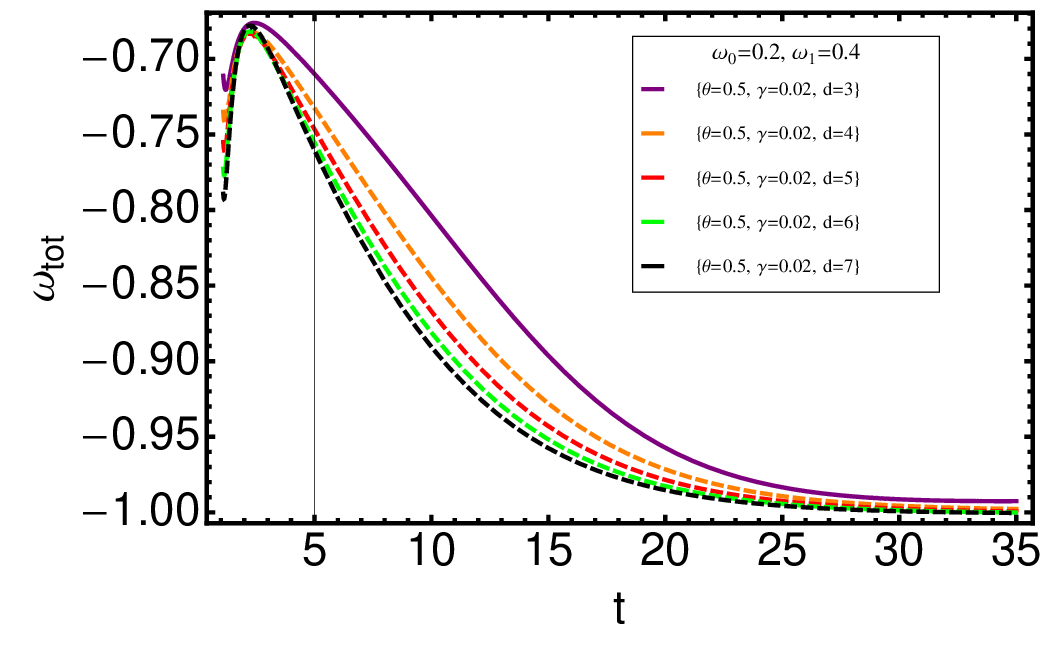}&\includegraphics[width=75 mm]{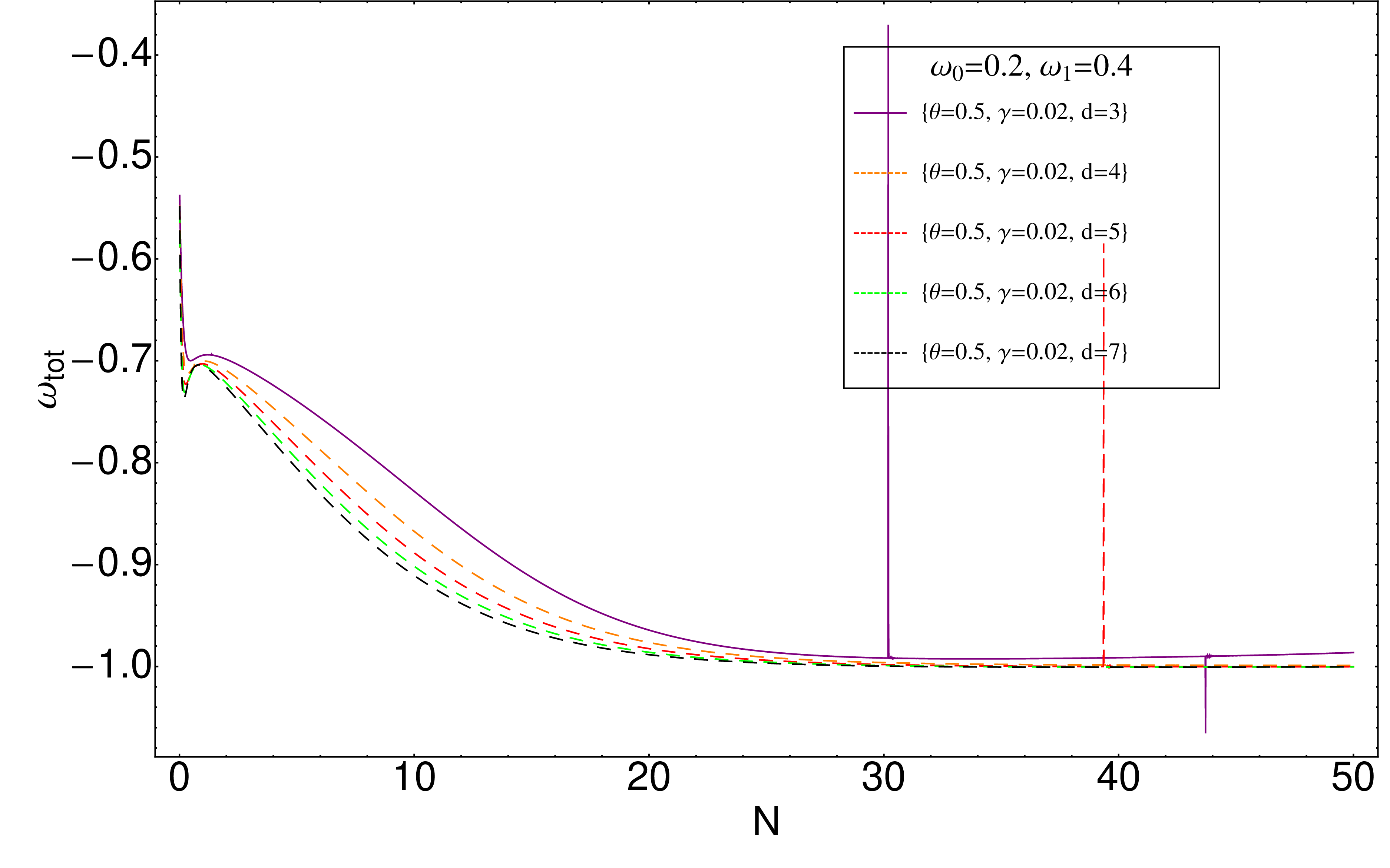}
 \end{array}$
 \end{center}
\caption{Behavior of $H$, $q$ and $\omega_{tot}$ against $t$ (left) and $N=\ln{a}$ (right). Model 3}
 \label{fig:4}
\end{figure}

The model 3 assumes the Universe where a dark matter modeled as a barotropic fluid with $\omega(t)_{b}$ given by the equation (19) and ghost dark energy interact with dark matter via,
\begin{equation}\label{eq:NintQ}
Q=(3+d)\gamma H \frac{\rho_{b}\rho_{G}}{\rho_{b}+\rho_{\small{G}}}+(3+d)\beta \dot{q} \frac{\rho_{b}\rho_{G}}{\rho_{G}-\rho_{b}},
\end{equation}
where $q$ is a deceleration parameter and $Q$ has a term containing time derivative of $q$. Here, $\gamma$ and $\beta$ are interaction coefficients, both specify interaction rate.  Dependence of $Q$ from $\rho{b}$ and $\rho_{G}$ is nonlinear.\\
Fig. 4 shows that Hubble expansion parameter have similar behavior with the previous models but with lower value, therefore not recover current observational value of $H$. Although changing sign of deceleration parameter and correct interval of total EoS illustrated in the Fig. 4, but we can not recommend the model 3 as real model of our Universe (because of Hubble expansion behavior).

\subsection{Model 4}

\begin{figure}[h!]
 \begin{center}$
 \begin{array}{cccc}
\includegraphics[width=75 mm]{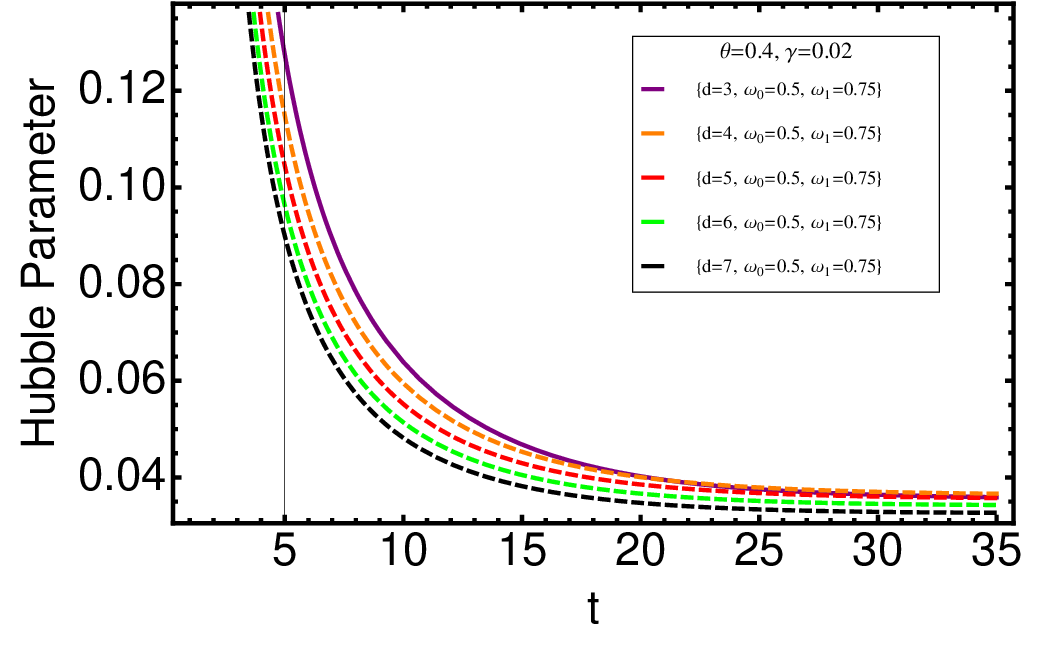} &
\includegraphics[width=75 mm]{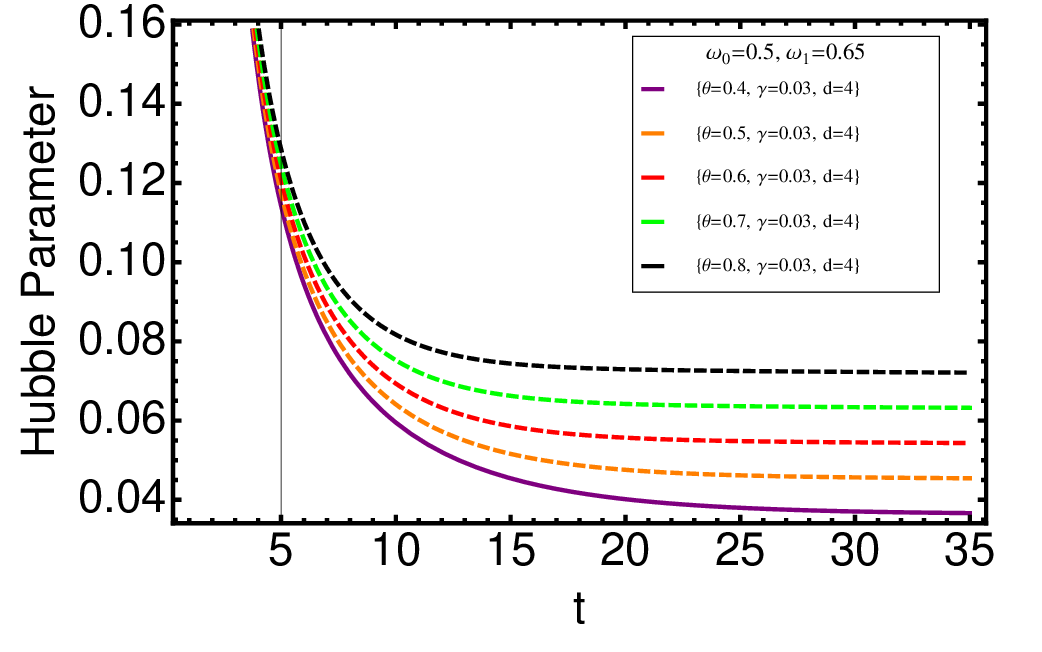}\\
\includegraphics[width=75 mm]{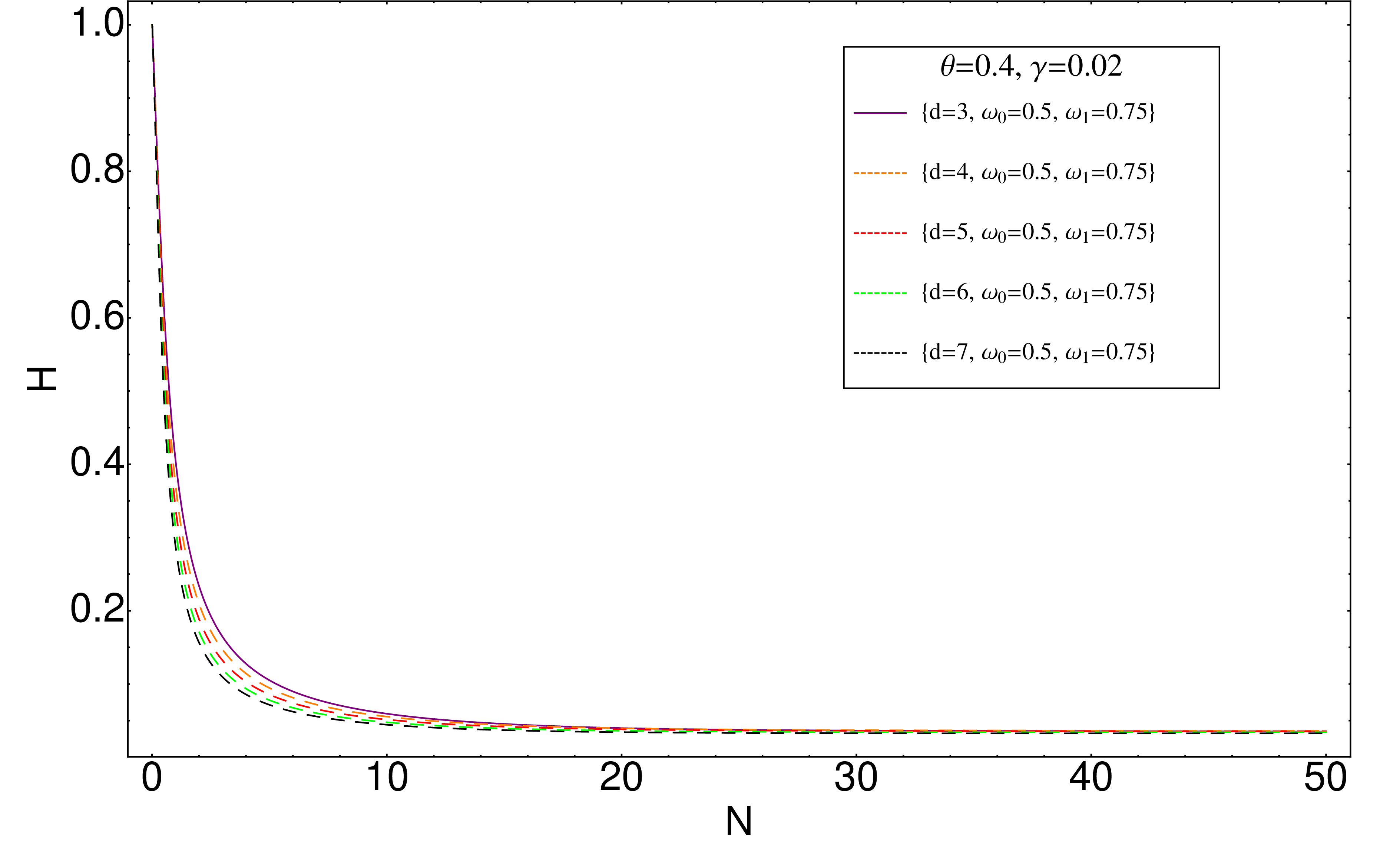} &
\includegraphics[width=75 mm]{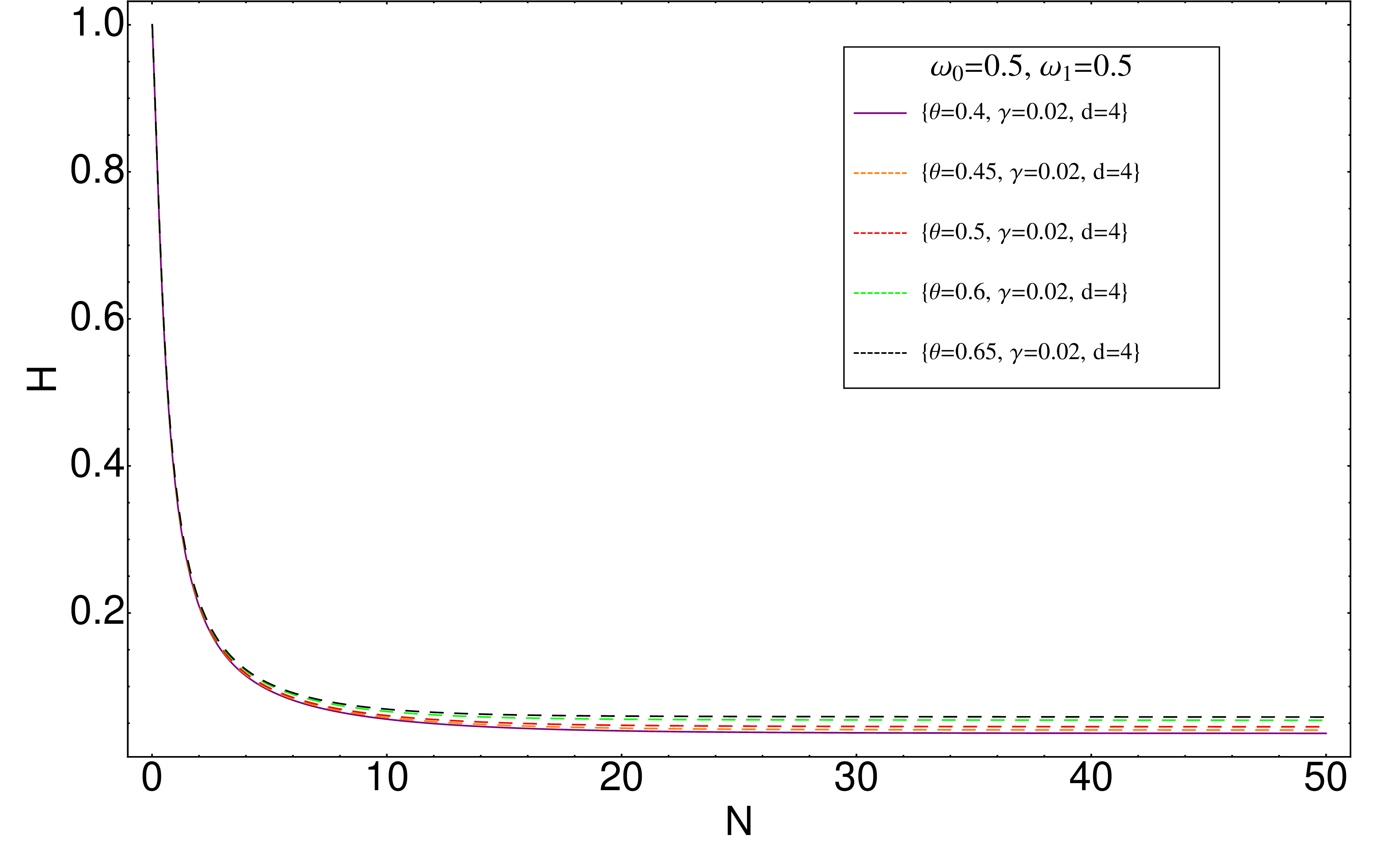}
 \end{array}$
 \end{center}
\caption{Behavior of $H$ against $t$ (up) and $N=\ln{a}$ (down). Model 4}
 \label{fig:5}
\end{figure}

\begin{figure}[h!]
 \begin{center}$
 \begin{array}{cccc}
\includegraphics[width=75 mm]{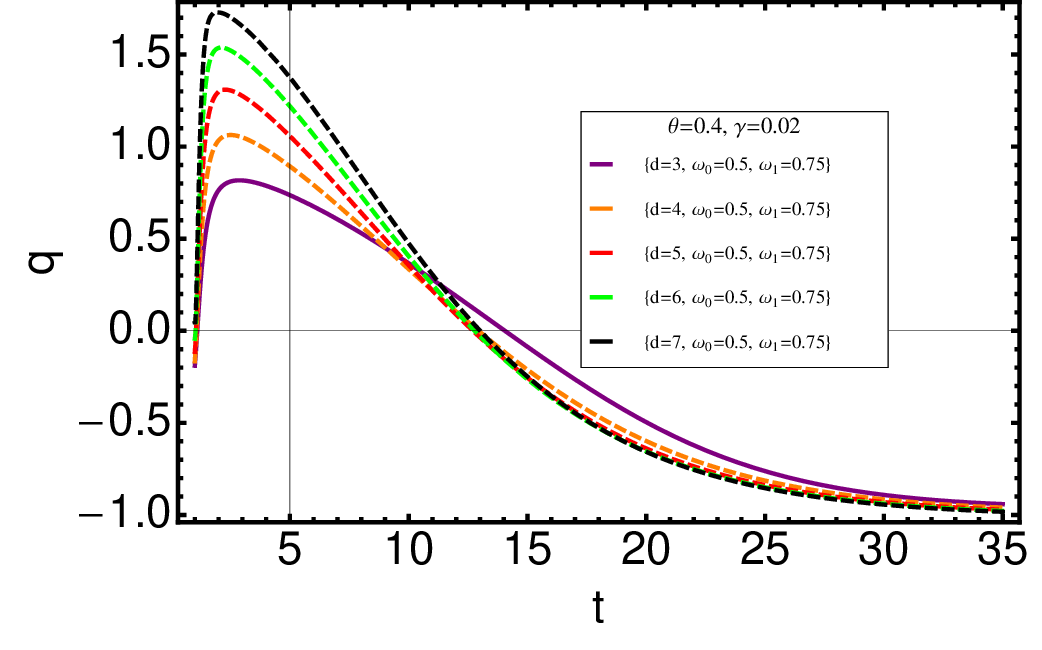}&
\includegraphics[width=75 mm]{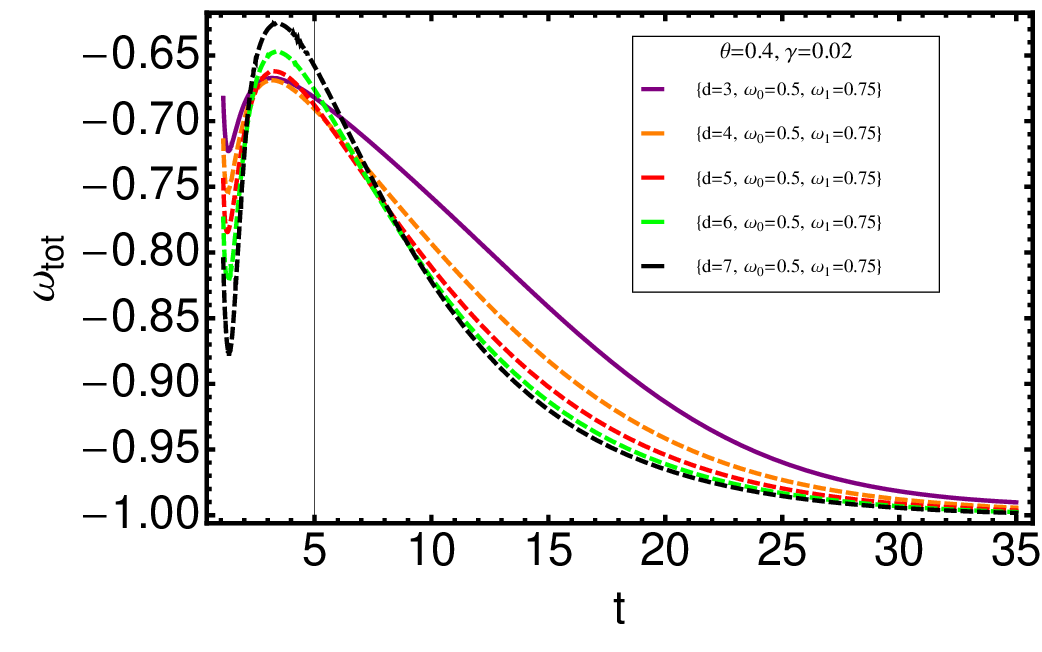}\\
 \end{array}$
 \end{center}
\caption{Behavior of $q$ and $\omega_{tot}$ against $t$ (up) and $N=\ln{a}$ (down). Model 4}
 \label{fig:6}
\end{figure}

For the model 4 we use $\omega(t)_{b}$ given by the equation (14) and interaction term $Q$ given by the equation (21). Our numerical results illustrated in the Figs. 5 and 6. In the Fig. 5 we can see variation of Hubble expansion parameter versus time which shows that is decreasing function of time and yields to a constant at the late time. Similar to the previous model the final constant value of $H$ does not match with current value of this parameter. As before we find that, increasing $d$ decreases value of Hubble expansion parameter. However the right plot shows that by fixing $\theta$ we can recover correct value for $H$ at the present epoch.\\
Then, Fig. 6 represent time evolution of deceleration parameter and total EoS parameter. We can see that increasing $d$ increases value of $q$ at the early stage and decreases one at the late stage. Also acceleration to deceleration phase transition represented. Then, we can see that total EoS parameter decreased suddenly at the early Universe and then grow up to a maximum, then decreased to reaches -1 at the late stage.

\section{Conclusion}
Recently, We studied interacting ghost dark energy models,
with variable $G$ and $\Lambda$ in 4 dimensions [48]. In that paper we obtained $\omega_{tot}\sim-0.3$. Now, in this paper we considered interacting ghost dark energy models with barotropic fluid in higher dimensional FRW space-time and extended Ref. [48] but with constant $G$ and $\Lambda$. One of the advantage of our new study is obtaining $\omega_{tot}\rightarrow-1$. We considered two different cases of barotropic EoS and two different cases of interaction terms. Therefore, we studied four different models and analyzed evolution of Hubble, deceleration and total EoS parameters numerically. We obtained effect of extra dimensions on these parameters and found that, extra dimensions decrease value of Hubble expansion parameter but effect of them on the deceleration and EoS parameters are depend on time. By comparing with observational data we concluded that the first and second models are more agree with current stage than the other models. Easily we can extend these models to the case of varying $G$ or $\Lambda$ to obtain more exact extension of the Ref. [48]. In fact, when we consider varying $G$ or $\Lambda$ of our models, can compare our results in higher dimensions with those obtained in [48] in four dimension.\\
In this paper, we used two class of interactions of the following general forms,
\begin{equation}\label{C1}
Q=\gamma H \rho,
\end{equation}
and
\begin{equation}\label{C2}
Q=\gamma H \rho+\beta \dot{q} \rho.
\end{equation}
while there are some different choices as [52],
\begin{equation}\label{C3}
Q\propto \dot{\rho},
\end{equation}
which may be considered in future works. Also it is interesting to apply current
observations, like CMB, BAO, $H_0$, and SNe, to Markov Chain Monte Carlo method to constrain the
model parameters and compare results with our paper.


\begin{thebibliography}{9}
\bibitem{Riess1}
A.G. Riess et al. [Supernova Search Team Colloboration], Astron. J.
116 (1998) 1009
\bibitem{Perlmutter2}
S. Perlmutter et al. [Supernova Cosmology Project Collaboration],
Astrophys. J. 517 (1999) 565
\bibitem{Amanullah3}
R. Amanullah et al., Astrophys. J. 716 (2010) 712
\bibitem{Pope4}
A.C. Pope et al. Astrophys. J. 607 (2004) 655,
[arXiv:astro-ph/0401249]
\bibitem{Spergel5}
D.N. Spergel et al. Astrophys. J. Supp. 148 (2003) 175,
[arXiv:astro-ph/0302209]
\bibitem{Sola6}
J. Sola and H. Stefancic, Phys. Lett. B 624 (2005) 147
\bibitem{Sola7}
I.L. Shapiro and J. Sola, Phys. Lett. B 682 (2009) 105
\bibitem{Ratra8}
B. Ratra and P. J. E. Peebles, Phys. Rev. D 37 (1988) 3406
\bibitem{Zlatev9}
I. Zlatev, L. M. Wang and P. J. Steinhardt, Phys. Rev. Lett. 82
(1999) 896
\bibitem{Saridakis10}
E.N. Saridakis and S. V. Sushkov, Phys. Rev. D 81 (2010) 083510
\bibitem{Singh11}
P. Singh, M. Sami and N. Dadhich, Phys. Rev. D 68 (2003) 023522
\bibitem{Cline12}
J.M. Cline, S. Jeon and G.D. Moore, Phys. Rev. D 70 (2004) 043543
\bibitem{Onemli13}
V.K. Onemli and R.P. Woodard, Phys. Rev. D 70 (2004) 107301
\bibitem{Hu14}
W. Hu, Phys. Rev. D 71 (2005) 047301
\bibitem{Setare15}
M.R. Setare and E. N. Saridakis, JCAP 0903 (2009) 002
\bibitem{Elizalde16}
E. Elizalde, S. Nojiri and S.D. Odintsov, Phys. Rev. D 70 (2004)
043539
\bibitem{Li17}
M.-Z Li, B. Feng, X.-M Zhang, JCAP, 0512 (2005) 002
\bibitem{Zhao18}
W. Zhao and Y. Zhang, Phys. Rev. D 73 (2006) 123509
\bibitem{Setare19}
M.R. Setare and E.N. Saridakis, Int. J. Mod. Phys. D 18 (2009) 549
\bibitem{Cai20}
Y. F. Cai, E. N. Saridakis, M. R. Setare and J. Q. Xia, Phys. Rept.
493 (2010) 1
\bibitem{Zhang21}
X. Zhang and F.Q. Wu, Phys. Rev. D 72 (2005) 043524
\bibitem{Nojiri22}
J. Sadeghi, B. Pourhassan, and Z. Abbaspour Moghaddam, "Interacting Entropy-Corrected Holographic Dark Energy and IR Cut-Off Length", IJTP 53 (2014) 125 [arXiv:1306.2055 [gr-qc]\
\bibitem{Elizalde23}
E. Elizalde, S. Nojiri, S.D. Odintsov and P. Wang, Phys. Rev. D 71
(2005) 103504
\bibitem{Li24}
H. Li, Z.K. Guo and Y.Z. Zhang, Int. J. Mod. Phys. D 15 (2006) 869
\bibitem{Saridakis25}
E.N. Saridakis, JCAP 0804 (2008) 020
\bibitem{Cai26}
R.G. Cai, Phys. Lett. B 657 (2007) 228
\bibitem{Wei27}
H. Wei and R.G. Cai, Phys. Lett. B 660 (2008) 113
\bibitem{Wei28}
H. Wei and R.G. Cai, Eur. Phys. J. C 59 (2009) 99
\bibitem{P29}
A.R. Amani and  B. Pourhassan, "Viscous Generalized Chaplygin gas
with Arbitrary $\alpha$", Int. J. Theor. Phys. 52 (2013) 1309
\bibitem{P30}
H. Saadat and  B. Pourhassan, "Viscous Varying Generalized Chaplygin
Gas with Cosmological Constant and Space Curvature", Int. J. Theor.
Phys. 52  (2013) 3712
\bibitem{P31}
H. Saadat and  B. Pourhassan, "FRW Bulk Viscous Cosmology with
Modified Chaplygin Gas in Flat Space", Astrophysics and Space
Science 343 (2013) 783
\bibitem{P32}
H. Saadat and  B. Pourhassan, "FRW bulk viscous cosmology with
modified cosmic Chaplygin gas", Astrophysics and Space Science 344
(2013) 237
\bibitem{P33}
B. Pourhassan, "Viscous Modified Cosmic Chaplygin Gas Cosmology"
International Journal of Modern Physics D 22 (9) (2013) 1350061
[arXiv:1305.6054 [gr-qc]]
\bibitem{P34}
J. Sadeghi, M. Khurshudyan, H. Farahani, "Phenomenological Varying
Modified Chaplygin Gas with Variable $G$ and $\Lambda$: Toy Models
for Our Universe", [arXiv:1308.1819 [gr-qc]]
\bibitem{Martiros35}
M. Khurshudyan, "Interaction between Generalized Varying Chaplygin
gas and Tachyonic Fluid", [arXiv:1301.1021 [gr-qc]]
\bibitem{Zhang36}
X. Zhang et al., JCAP 0601 (2006) 003
\bibitem{Chattopadhyay37}
S. Chattopadhyay, U. Debnath, Grav. Cosmol. 14 (2008) 341
\bibitem{Jamil38}
M. Jamil, "Interacting New Generalized Chaplygin Gas", Int. J.
Theor. Phys. 49 (2010) 62
\bibitem{Jamil39}
H. Saadat and  B. Pourhassan "Effect of Varying Bulk Viscosity on Generalized Chaplygin Gas", IJTP DOI 10.1007/s10773-013-1913-8
\bibitem{40}
J. Sadeghi, B. Pourhassan, M. Khurshudyan, H. Farahani, "Time-Dependent Density of Modified Cosmic Chaplygin Gas with Cosmological Constant in Non-Flat Universe" IJTP DOI 10.1007/s10773-013-1881-z
\bibitem{41}
F.R. Urban, A.R. Zhitnitsky, Phys. Rev. D 80 (2009) 063001
\bibitem{42}
F.R. Urban, A.R. Zhitnitsky, JCAP 09 (2009) 018
\bibitem{43}
F.R. Urban, A.R. Zhitnitsky, Nucl. Phys. B 835 (2010) 135
\bibitem{44}
E. Ebrahimi, A. Sheykhi, Int. J. Mod. Phys. D 20 (2011) 2369
\bibitem{Chao-Jun45}
Chao-Jun Feng, Xin-Zhou Li, Ping Xi, "Global behavior of
cosmological dynamics with interacting Veneziano ghost", JHEP 1205
(2012) 046
\bibitem{46}
Chao-Jun Feng, Xin-Zhou Li, Xian-Yong Shen, "Latest Observational
Constraints to the Ghost Dark Energy Model by Using Markov Chain
Monte Carlo Approach", Phys. Rev. D87 (2013) 023006
\bibitem{AmalyaMartiros47}
M. Khurshudyan, A. Khurshudyan, "A model of a varying Ghost Dark
energy", [arXiv:1307.7859[gr-qc]]
\bibitem{48}
J. Sadeghi, M. Khurshudyan, A. Movsisyan and H. Farahan, "Interacting ghost dark energy models
with variable G and $\Lambda$", JCAP12(2013)031
\bibitem{49}
S. Chakraborty, U. Debnath, M. Jamil, "Variable $G$ Correction for Dark Energy Model in Higher Dimensional Cosmology", Canadian Journal of Physics 90 (2012) 365
\bibitem{50}
A.R. Amani, B. Pourhassan, "FRW Cosmology and Static Extra Dimension
with Non-zero Cosmological Constant", Int. J. Theor. Phys. 51 (2012) 49
\bibitem{51}
I. Pahwa, D. Choudhury, T.R. Seshadri, "Late-time acceleration in Higher Dimensional Cosmology", JCAP 1109 (2011) 015
\bibitem{52}
M. Shahalam, S. D. Pathak, M.M. Verma, M. Yu. Khlopov, R. Myrzakulov,  Eur Phys. Jour. C
75 (2015) 395
\end{thebibliography}
\end{document}